\documentclass[aps,preprint,showpacs,showkeys]{revtex4}   
\usepackage{epsfig}  
\usepackage{ amsmath,amssymb,amsthm}    
\usepackage{graphicx}
\usepackage{psfrag}
\usepackage{bm}
\usepackage[dvips]{color}   
\newcounter{egyenlet}[equation]   
   
\begin{document}   
   
\renewcommand{\thesection}{\arabic{section}}   
\renewcommand{\theequation}{\arabic{equation}\alph{egyenlet}}   
\def\dl{\displaystyle}    
\def\arrowlimit#1{\mathrel{\mathop{\longrightarrow}\limits_{#1}}}   
\def\grad{{\rm grad}}   
\def\div{{\rm div}}   
\def\rot{{\rm rot}}   
\def\scs{\scriptsize}   
\def\l{\langle}   
\def\r{\rangle}

\title{\Large A Generalized Statistical Complexity Measure: \\
Applications to Quantum  Systems}   
   
\author{\bf R. L\'opez-Ruiz$^1$, \'A. Nagy$^2$, E. Romera$^3$ and J. Sa\~nudo$^4$}  
\affiliation{  
\small$^1$DIIS and BIFI, Facultad de Ciencias, Universidad de Zaragoza, 
   E-50009 Zaragoza, Spain \\ 
$^2$Department of Theoretical Physics, University of Debrecen, 
   H-4010 Debrecen, Hungary\\   
$^3$Departamento de F\'{\i}sica At\'omica, Molecular y Nuclear, 
and Instituto  Carlos I de F{\'\i}sica Te\'orica y Computacional,   
Universidad de Granada, E-18071 Granada, Spain \\   
$^4$Departamento de F\'{\i}sica, Facultad de Ciencias, Universidad de Extremadura,   
E-06071 Badajoz, Spain,
and BIFI, Universidad de Zaragoza, E-50009 Zaragoza, Spain}  
 
\date{\today}

\begin{abstract}   
A two-parameter family of complexity measures $\tilde{C}^{(\alpha,\beta)}$ 
based on the R\'enyi entropies is introduced and 
characterized by a detailed study of its mathematical properties. 
This family is the generalization of a continuous version of the LMC complexity,
which is recovered for $\alpha=1$ and $\beta=2$.
These complexity measures are obtained by multiplying two quantities
bringing global information on the probability distribution defining the system.
When one of the parameters, $\alpha$ or $\beta$, goes to infinity, 
one of the global factors becomes  a local factor. For this special case,
the complexity is calculated on different quantum systems: 
H-atom, harmonic oscillator and square well.  
\end{abstract}  
  

\maketitle

\section{Introduction} 
 
The study of statistical measures in physical systems, and in particular in 
quantum systems, has a role of growing importance. So, information entropies 
and statistical  complexities have been calculated on different atomic systems 
\cite{gadre1985,panos2005}. In particular, the so called 
LMC  complexity \cite{lopez1995,catalan2002} has been 
computed in the position and momentum spaces for the density functions of 
the Hydrogen-like atoms and the quantum isotropic harmonic oscillator \cite{sanudo2008,sanudo2008+}.
It has been found that the minimum values of that statistical measure is taken on the   
quantum states with the highest orbital angular momentum, 
just those wave functions that correspond to the Bohr-like orbits in the pre-quantum 
image. 
 
Many LMC-like statistical complexities are defined as a product of two factors, one of them 
measuring the broadening of the distribution that defines the system and the other one 
quantifying the narrowness of it. Both factors are global magnitudes that can be 
calculated by integrating over the whole support of the distribution.

Shannon information \cite{shannon1948} is an adequate indicator to grasp the spreading of a distribution
and thus it is employed as a basic ingredient of the first factor of complexity measures.
Concretely, it plays an important role in the original LMC statistical complexity
in which the second factor, the so called disequilibrium \cite{lopez1995}, 
is the square distance to the equiprobability distribution. 
Different generalizations of the Shannon information that depend on  
a parameter can be found in the literature \cite{renyi1961,tsallis1988,rosso2006}.  
For instance, the R\'enyi entropy, that can be related with relevant physical magnitudes in
atomic physics \cite{liu1,liu2}, is a good candidate to be used in this purpose.

Hence, in this work, we undertake the generalization of the second factor of the LMC complexity
by means of the R\'enyi entropy. We apply the same procedure than in \cite{arxiv2009} where
the first factor was generalized. This is presented in Section 2. 
Some properties of this new two-parameter-dependent 
complexity measure are indicated in Section 3. 
Strikingly, when one of the parameters tends to infinity,
the asymptotic limit of this measure becomes the product of a global quantity by a local one. 
In Section 4, the analysis and calculation of this special global/local product case of the generalized 
complexity measure are performed for the H-atom, the quantum harmonic oscillator and the square well. 
Last section includes the conclusions.

\section{Generalized Statistical Complexity Measure $\tilde{C}_f^{(\alpha,\beta)}$}  

Let us consider a $D$-dimensional density function $f({\bf r})$, (with  
$f({\bf r})$ nonnegative and {$\int f({\bf r}) d{\bf r}=1$).
The R\'enyi entropy of order $\alpha$ of the density function $f$ is given by  
\begin{equation}  
R_f^{(\alpha)} = \frac{1}{1-\alpha}\ln\int [f({\bf r})]^{\alpha} d{\bf r}, 
\quad \text{for } 0<\alpha<\infty, \,\, \alpha\neq 1, 
\end{equation}  
where $\bf r$ stands for $r_1, ..., r_D$. From the above definition, it is 
straightforward to see that in the limit $\alpha\rightarrow 1$  we have 
$R_{f}^{(\alpha)}\rightarrow S_f = -\int f({\bf r})\ln f({\bf r})d{\bf r}$, 
with $S_f$ the Shannon entropy of $f$, and in the limit $\alpha\rightarrow\infty$ then
$R_f^{(\alpha)}\rightarrow -\ln ||f||_{\infty}$, where $||f||_{\infty}=sup_{\bf r} f({\bf r})$
represents the maximum reached by $f$ over its whole support. 

The importance of R\'enyi entropies comes from the fact that,
for atoms and molecules, density functionals (kinetic energy, exchange energy and 
classical Coulomb repulsion energy)
can be expanded in terms of the local homogeneous functionals $\exp{(1-\alpha)
R_{\rho}^{(\alpha)}}$ \cite{liu1,liu2}. In particular, it is well known that
for $\alpha=5/3$, $\alpha=4/3$ and $\alpha=2$, the local density approximations are 
related to the kinetic and exchange energies and the average of the density, respectively.

A continuous version \cite{catalan2002} of the measure of complexity $C_{f}$,
the so-called LMC complexity introduced in \cite{lopez1995}, is defined by   
\begin{equation}  
C_f=H_fQ_f, \quad \text{with} \quad H_f=e^{S_f}\quad \text{and} \quad Q_f=e^{-R_f^{(2)}}.  
\end{equation}  
When the Shannon entropy  of the statistical complexity $C_f$ is replaced with 
the R\'enyi entropy of order $\alpha$, we obtain the generalized statistical 
measure of complexity, $C_f^{(\alpha)}$, which has been defined by \cite{arxiv2009}  
\begin{equation}  
C_f^{(\alpha)}=H_f^{(\alpha)}Q_f, \quad \text{with} \quad H_f^{(\alpha)} = 
e^{R_f^{(\alpha)}},  
\end{equation}  
and tends to $C_f$ in the limit $\alpha\rightarrow 1$. 
 
Now we can substitute in a symmetric way the $R_f^{(2)}$ ingredient of the above complexities for the R\'enyi 
entropy of order $\beta$, which allows us to obtain a 
$(\alpha,\beta)-$dependent measure of complexity, 
$\tilde{C}^{(\alpha,\beta)}_f$, which is defined by 
\begin{equation} 
\tilde{C}^{(\alpha,\beta)}_f=e^{R_f^{(\alpha)}-R_f^{(\beta)}},\quad \quad 0<\alpha,\beta<\infty\,.  
\label{comp} 
\end{equation}  
So,  we recover $\tilde{C}^{(1,2)}_f=C_f$ and 
$\tilde{C}^{(\alpha,2)}_f=C_f^{(\alpha)}$ of Refs. \cite{lopez1995,arxiv2009}, respectively.
This type of generalization based on R\'enyi entropies differences was suggested in \cite{lopezruiz2005}
after the work of Varga and Pipek \cite{varga2003} in this same line of thought.

\section{Properties of $\tilde{C}_f^{(\alpha,\beta)}$ }

Now, having as guideline Ref. \cite{catalan2002}, 
we proceed to present some mathematical properties 
of this new generalized statistical complexity measure.

\subsection{Inversion symmetry} 

It is straightforward to check that 
\begin{equation}
\tilde{C}^{(\alpha,\beta)}_f \tilde{C}^{(\beta,\alpha)}_f=1,
\end{equation}
and then $\tilde{C}^{(\alpha,\alpha)}_f=1$.

\subsection{Monotonicity and universal bound}

Taking into account that the R\'enyi entropy is a nonincreasing function of $\alpha$,
it can be easily proved that  
\begin{itemize} 
\item[(i)] $\tilde{C}_f^{(\alpha,\beta)}\geq 1$ if $\alpha<\beta$ and
  $\tilde{C}_f^{(\alpha,\beta)}\leq 1$ if $\alpha>\beta$. 
\item[(ii)]  $\tilde{C}_f^{(\alpha,\beta)}$ is a nonincreasing function of $\alpha$ for a 
  fixed $\beta$ and an increasing function of $\beta$ for a fixed $\alpha$. 
\item[(iii)] The lower (upper) bound 1 is reached for $\alpha<\beta$ 
($\alpha>\beta$) for the rectangular density function. It is a universal bound independent
of $\alpha$ and $\beta$, as it is shown in Sec. \ref{ssf}. 
\end{itemize}

\subsection{Invariance under translations and rescaling transformations} 

$\tilde{C}_f^{(\alpha,\beta)}$ is invariant under scaling transformations, i. e. for 
$f_{\lambda}=\lambda^{D}f(\lambda {\bf r})$, then 
$\tilde{C}_{f_{\lambda}}^{(\alpha,\beta)}=\tilde{C}_f^{(\alpha,\beta)}$.
Also, it is invariant under translations. Hence, in general, let us consider a scaling 
transformation and a translation parameterized by $(a,{\bf b})$, respectively, of the form 
\begin{equation} 
\label{a1} 
f_{a{\bf b}}({\bf r})=a^D f[a({\bf r}-{\bf b})] , 
\end{equation} 
where the distribution function $f$ is normalized, $\int f({\bf r})d{\bf r} = 1$. 

The R\'enyi entropy of order $\alpha$ is transforming as 
\begin{equation} 
\label{a3} 
R_{a{\bf b}}^{(\alpha)}= \frac{1}{1-\alpha} \ln{\int (a^D f[a({\bf r}-{\bf b})])^{\alpha}} d{\bf r} 
= R^{(\alpha)} - D \ln{a} , 
\end{equation} 
where a change of variable ${\bf y}=a({\bf r}-{\bf b})$ was applied. 
Note that there is no dependence on the parameter ${\bf b}$. Therefore 
the new complexity measure $\tilde{C}_f^{(\alpha,\beta)}$ (\ref{comp}) is invariant 
under this transformation. 
 
\subsection{Invariance under replication} 
 
Take $n$ copies $f_m({\bf r}), m=1, ..., n$ of the distribution function  
$f({\bf r})$,
\begin{equation} 
\label{b1} 
f_{m}({\bf r})=n^{D/2-1} f[n^{1/2}({\bf r}-{\bf b}_m)] , \quad 1 \le m \le n, 
\end{equation} 
where the support of each $f_m({\bf r})$ is centered at the point ${\bf b}_m$ and 
the supports are disjoint. We can immediately obtain that  
$\int f_{m}({\bf r})d{\bf r} = 1/n$. Then we can define the distribution function \cite{catalan2002}
\begin{equation} 
\label{b3} 
 q({\bf r}) = \sum_{m=1}^{n}f_{m}({\bf r})  , 
\end{equation} 
that is normalized, $\int q({\bf r})d{\bf r} = 1$. 
From Eq. (\ref{b1}) we can easily calculate that  
\begin{equation} 
\label{b5} 
\int f_{m}^{\alpha}({\bf r}) d{\bf r} =n^{(\alpha-1)D/2-\alpha} \int 
f^{\alpha}({\bf y})d{\bf y} \,.
\end{equation} 
As the replicas are supported on disjoint sets, we have 
\begin{equation} 
\label{b6} 
\sum_{m=1}^{n} \int f_{m}^{\alpha}({\bf r}) d{\bf r} = 
n^{(\alpha-1)(D/2-1)} \int 
f^{\alpha}({\bf y})d{\bf y} \,.
\end{equation} 
Then the transformation of the R\'enyi entropy of order $\alpha$ is 
\begin{equation} 
\label{b7} 
R_{q}^{(\alpha)}= R_f^{(\alpha)} - (D/2-1) \ln{n} \,. 
\end{equation} 
Hence the  complexity measure $\tilde{C}_q^{(\alpha,\beta)}$ (\ref{comp}) 
is replica invariant. 
 
\subsection{Near-continuity} 
 
Take two distribution functions $f({\bf r})$ and $g({\bf r})$ defined
on the set $M$ in the $D$ dimensional space considered. Let $\delta$ be a
positive real number. The   functions $f({\bf r})$ and $g({\bf r})$
are $\delta$-neighboring functions on  $M$, if the Lebesgue measure of
the points ${\bf r} \in M $ satisfying 
$|f({\bf r})-g({\bf r})| \ge \delta$ is zero. A functional $T$ of the
distribution functions is near-continuous if for any $\varepsilon >0$ 
there exists $\delta(\varepsilon) >0$ such that for any $\delta$-neighboring 
functions $f({\bf r})$ and $g({\bf r})$ on $T$ then $|T(f)-T(g)|<\varepsilon$. 
 
Take the function 
\begin{equation} 
g_{\delta,B}({\bf r}) 
= 
\left\{ 
\begin{array}{ll}
\frac{1- \delta}{c_D}  & if~ |{\bf r}|<1
\\
\frac{\delta}{c_D (B^D - 1)}    & if~1 < |{\bf r}| < B \\
0 & ~ \text{otherwise}
\end{array}
\right. \; ,
\label{c1}
\end{equation}
where 
\begin{equation}
\label{c3}
c_D = \frac{2 \pi^{D/2}}{D \Gamma(D/2)} \, ,
\end{equation}
where $B > 1$ and $1 > \delta > 0$. As $c_D$ is the volume of a unity $D$-dimensional
sphere then $g$ is normalized to 1.
One can easily calculate the R\'enyi entropy of order $\alpha$,
\begin{equation}
\label{c4}
R_{g}^{(\alpha)}= \frac{1}{1-\alpha} \ln{ \left[(1-\delta)^{\alpha} +
\frac{\delta^{\alpha}}{(B^D -1)^{\alpha-1}}\right]} +\ln{c_D} \, .
\end{equation}
From here, we obtain $\tilde{C}^{(\alpha,\beta)}_g=e^{R_g^{(\alpha)}-R_g^{(\beta)}},
0<\alpha,\beta<\infty$. Consider now the rectangular density function 
\begin{equation} 
\chi({\bf r}) =  
\left\{ 
\begin{array}{ll} 
\frac{1}{c_D}  & if~ |{\bf r}|<1  \\ 
  0  & ~ \text{otherwise} 
\end{array} 
\right. \, .
\label{c6} 
\end{equation} 
The R\'enyi entropy of order $\alpha$  
\begin{equation} 
\label{c7} 
R_{\chi}^{(\alpha)}= \ln{c_D}  
\end{equation} 
does not depend on the $\alpha$. Consequently, we are led to the result  
\begin{equation} 
\label{c17} 
\tilde{C}_{\chi}^{(\alpha,\beta)}= 1  \, . 
\end{equation} 
Note that $g_{\delta,B}$ and $\chi$ are $\tilde{\delta}$-neighboring functions 
for $0 < \delta < \tilde{\delta} < 1$ and  
\begin{equation} 
\label{c8} 
lim_{{\delta} \to 0} R_{g}^{(\alpha)} = R_{\chi}^{(\alpha)}= \ln{c_D} \, . 
\end{equation} 
Therefore in the limit ${\delta} \to 0$ the complexity measure  
$\tilde{C}^{(\alpha,\beta)}$ takes the same value:  
\begin{equation} 
\label{c9} 
lim_{{\delta} \to 0} \tilde{C}^{(\alpha,\beta)}_g =  
\tilde{C}^{(\alpha,\beta)}_{\chi} = 1 . 
\end{equation} 
The importance of a bounded support to obtain this result deserves
some longer explanation to be done in a future work,
such as it was suggested in \cite{catalan2002}.

\subsection{The extremal complexity} 
\label{ssf}

The extremal complexity is reached for the rectangular function  
 (Eqs. (\ref{c6}) and  (\ref{c17})). It can be proved following  
\cite{catalan2002}. The function $f$ is taken as a sum of rectangular functions 
$\chi_{E_k}$ defined on disjoint sets $E_k, k=1,...,n$ with Lebesgue measure  
$\mu_k$ 
\begin{equation} 
\label{d1} 
f({\bf r}) = \sum_{k=1}^{n} \lambda_k \chi_{E_k}  \, . 
\end{equation} 
Its integrals can be easily calculated: 
\begin{equation} 
\label{d2} 
\int f^{\alpha}d{\bf r} = \sum_{k=1}^{n} \lambda_k^{\alpha} \mu_k  \, . 
\end{equation} 
The logarithm of complexity measure $\tilde{C}^{(\alpha,\beta)}$ has the form 
\begin{equation} 
\label{d3} 
\ln{\tilde{C}^{(\alpha,\beta)}} =  
\frac{1}{1-\alpha}\ln{\left(\sum_{k=1}^{n} \lambda_k^{\alpha} \mu_k\right)} -  
\frac{1}{1-\beta}\ln{\left(\sum_{k=1}^{n} \lambda_k^{\beta} \mu_k\right)}  \, . 
\end{equation} 
We seek the extremum of the logarithm of complexity measure  
$\tilde{C}^{(\alpha,\beta)}$ under the normalization condition  
\begin{equation} 
\label{d4} 
\sum_{k=1}^{n} \lambda_k \mu_k = 1 \, . 
\end{equation} 
The variation with respect to $\lambda_k$ and $\mu_k$ after straighforward  
manipulation leads the equations 
\begin{equation} 
\label{d5} 
\lambda_k^{\alpha-\beta} = \frac{\sum_{l=1}^{n} \lambda_l^{\alpha} \mu_l} 
{\sum_{l=1}^{n} \lambda_l^{\beta} \mu_l}   
\end{equation} 
for all $k=1,...,n$. As $\lambda_k$ has the same value for all $k$, then $f$ is a 
rectangular function. Moreover, with the help of the near-continuity property as explained
in \cite{catalan2002}, it can also be argued that the rectangular distribution is the only 
distribution reaching the extremal complexity. 

\subsection{The cases $\tilde{C}_f^{(\alpha,\infty)}$ and $\tilde{C}_f^{(\alpha,0)}$} 

As before explained,
the new complexity measure has a completely different behavior for  
$\alpha<\beta$ and $\alpha>\beta$. In the first case, there is a lower bound, 
and in the second case there is an upper bound. Both are universal, i. e. the  
bound is equal to $1$ for any (finite and non zero) value of $\alpha$ or  
$\beta$, and this universal bound is reached at the rectangular density  
function. This is a consequence of the fact that the 
R\'enyi entropy is independent of the parameter $\alpha$ or  
$\beta$, then the new complexity measure has the value of $1$ for any 
 $\alpha$ or $\beta$. 
 
Finally, let us remark that when $\beta$ goes to infinity   
a  special case of the complexity measure in terms of a {\em local quantity} ($||f||_{\infty}= 
sup_{\bf r}f({\bf r}))$ is obtained: 
$\tilde{C}_f^{(\alpha,\infty)}=e^{R_f^{(\alpha)}} ||f||_{\infty}$. 
To prove this property  it is sufficient to take into account that $\lim_{p\rightarrow\infty} 
\left(\int f({\bf r})^p d{\bf r}\right)^{1/p}=sup_{\bf r}f({\bf r})$ \cite{debnath}.
This complexity measure verifies 
$\tilde{C}_f^{(\alpha,\infty)}>1$ for all $\alpha$ (and $\tilde{C}_f^{(\infty,\alpha)}<1$). 
We can also mention that $\tilde{C}_f^{(\alpha,0)} \to 0$  for finite $\alpha$ and  
$\beta \to 0$.

\section{Calculation of $\tilde{C}_f^{(\alpha,\infty)}$ on different quantum systems}   
  
Among the different  statistical indicators that have been defined as a  
product of two entropic terms, 
the generalized complexity $\tilde{C}^{(\alpha,\beta)}$ is also obtained by multiplying two factors, 
each one of them bringing global information on the distribution $f$.   
Recall, however, that the limit $\tilde{C}_f^{(\alpha,\infty)}$  
combines global information on the distribution $f$, just the part corresponding to $e^{R_f^{(\alpha)}}$,  
and local information coming from only a specific point of the space where the distribution $f$ is supported,  
indeed the maximum of the density, $||f||_{\infty}$. Equivalently, the limit version $\tilde{C}_f^{(\infty,\beta)}$ 
presents a symmetrical behavior respect to $\tilde{C}_f^{(\alpha,\infty)}$ 
by considering the property $\tilde{C}_f^{(\infty,\sigma)}\tilde{C}_f^{(\sigma,\infty)}=1$.  
Here we undertake the calculation of $\tilde{C}_f^{(\alpha,\infty)}$ for three paradigmatic quantum systems: 
the $H$-atom, the harmonic oscillator and the square well. 
   
\subsection{$H$-Atom} 
 
The atomic state of the $H$-atom determined by the quantum numbers $(n,l,m)$ 
in position space ($\vec{r}=(r,\Omega)$, with $r$ the radial distance and $\Omega$ the solid angle)  
is given by the non-relativistic wave function 
\begin{equation} 
\Psi_{n,l,m}(\vec{r})= R_{n,l}(r)\; Y_{l,m}(\Omega)\;, 
\end{equation} 
where $R_{n,l}(r)$ is the radial part and $Y_{l,m}(\Omega)$ is the spherical harmonic.  
The radial part is expressed  
as \cite{galindo1991} 
\begin{equation} 
R_{n,l}(r)= {2\over n^2} \left[{(n-l-1)!\over (n+l)!}\right]^{1/2}\; 
\left({2r\over n}\right)^l\;e^{-{r\over n}}\; L_{n-l-1}^{2l+1}\left({2r\over n}\right)\;, 
\end{equation} 
 $L_{\alpha}^{\beta}(t)$ being  the associated Laguerre polynomials. 
Atomic units are used in this section. 
 
The same hydrogenic atomic state in momentum space ($\vec{p}=(p,\hat{\Omega})$,  
with the momentum modulus $p$ and the solid angle $\hat{\Omega}$) is given by  
the wave function 
\begin{equation} 
\hat{\Psi}_{n,l,m}(\vec{p})= \hat{R}_{n,l}(p)\; Y_{l,m}(\hat{\Omega})\;, 
\end{equation} 
where the radial part $\hat{R}_{n,l}(p)$ is expressed now as \cite{bethe1977} 
\begin{equation} 
\hat{R}_{n,l}(p)= \left[{2\over\pi}{(n-l-1)!\over (n+l)!}\right]^{1/2}\; 
n^2\;2^{2l+2}\;l!\;{n^lp^l\over (n^2p^2+1)^{l+2}}\;  
G_{n-l-1}^{l+1}\left({n^2p^2-1\over n^2p^2+1}\right)\;, 
\end{equation} 
with the Gegenbauer polynomials $G_{\alpha}^{\beta}(t)$. 
 
Taking the former expressions, the probability density 
in position and momentum spaces, 
\begin{equation} 
\rho(\vec{r})\;=\;\mid\Psi_{n,l,m}(\vec{r})\mid^2\;, \hspace{1cm} 
\gamma(\vec{p})\;=\;\mid\hat{\Psi}_{n,l,m}(\vec{p})\mid^2\;, 
\end{equation} 
can be explicitly calculated. From these densities, we compute $\tilde{C}_f^{(\alpha,\infty)}$, 
taking into account that for these cases $f=\rho(\vec{r})$ or $f=\gamma(\vec{p})$,  
respectively. 
 
In Figure 1, the value of the generalized complexity in position space,  
$\tilde{C}_{r}^{(\alpha,\infty)}$, is shown for   
$n=15$ and $l=5,10,14$ versus $|m|$ with $\alpha=0.5$ (Fig. 1(a)) and $\alpha=2$ (Fig. 1(b)). 
The same calculation in momentum space, $\tilde{C}_{p}^{(\alpha,\infty)}$, 
can be seen in Figs. 2(a) and 2(b), for the cases $\alpha=0.5$ and $\alpha=2$, respectively. 
Note that the minimum of the generalized complexity corresponds    
 just to the highest $l$ for a given $n$ in both position and momentum spaces. 
This property is independent of the parameter $\alpha$. 
   
\subsection{Harmonic oscillator} 
 
Let us consider a particle under the action of the potential energy  
$V(r)=\lambda^2r^2/2$, where $\lambda$ is a positive real constant expressing  
the potential strength. The three-dimensional non-relativistic wave functions  
of this system in position space ($\vec{r}=(r,\Omega)$) are: 
\begin{equation} 
\Psi_{n,l,m}(\vec{r})= R_{n,l}(r)\; Y_{l,m}(\Omega)\;, 
\label{eq:Psi} 
\end{equation} 
where $R_{n,l}(r)$ is the radial part and $Y_{l,m}(\Omega)$ is the spherical harmonic  
of the quantum state determined by the quantum numbers $(n,l,m)$. The radial part is expressed  
as \cite{galindo1991} 
\begin{equation} 
R_{n,l}(r)= \left[{2\;n!\;\lambda^{l+3/2}\over \Gamma(n+l+3/2)}\right]^{1/2}\; 
r^l\;e^{-{\lambda \over 2}r^2}\; L_{n}^{l+1/2}(\lambda r^2)\;, 
\label{eq:R} 
\end{equation} 
where $L_{\alpha}^{\beta}(t)$ are the associated Laguerre polynomials. 
The levels of energy are given by  
\begin{equation} 
E_{n,l}=\lambda (2n+l+3/2) = \lambda (e_{n,l}+3/2), 
\label{eq:E} 
\end{equation} 
where $n=0,1,2,\cdots$ and $l=0,1,2,\cdots$. 
Let us observe that $e_{n,l}=2n+l$. 
Thus, different pairs of $(n,l)$ can give the same $e_{n,l}$, 
and then the same energy $E_{n,l}$. 
 
The wave functions in momentum space ($\vec{p}=(p,\hat{\Omega})$) are: 
\begin{equation} 
\hat{\Psi}_{n,l,m}(\vec{p})= \hat{R}_{n,l}(p)\; Y_{l,m}(\hat{\Omega})\;, 
\label{eq:Psih} 
\end{equation} 
where the radial part $\hat{R}_{n,l}(p)$ is now given by the expression 
\begin{equation} 
\hat{R}_{n,l}(p)= \left[{2\;n!\;\lambda^{-l-3/2}\over \Gamma(n+l+3/2)}\right]^{1/2}\; 
p^l\;e^{-{p^2\over 2\lambda}}\; L_{n}^{l+1/2}({p^2/\lambda})\;. 
\label{eq:Rh} 
\end{equation} 
 
Taking the former expressions, the probability density 
in position and momentum spaces, 
\begin{equation} 
\rho_{\lambda}(\vec{r})\;=\;\mid\Psi_{n,l,m}(\vec{r})\mid^2\;, \hspace{1cm} 
\gamma_{\lambda}(\vec{p})\;=\;\mid\hat{\Psi}_{n,l,m}(\vec{p})\mid^2\;, 
\label{eq:rho} 
\end{equation} 
can be explicitly calculated. From these densities,  
the generalized statistical complexity is computed. 
The subindex $\lambda$ can be dropped because this indicator, $\tilde{C}^{(\alpha,\beta)}$, 
is independent of the potential strength, $\lambda$,  
due to its invariance under scaling transformation. 
As a consequence of this property, it is also found that this magnitude is the same 
in both position and momentum spaces, $\tilde{C}_r^{(\alpha,\beta)}=\tilde{C}_p^{(\alpha,\beta)}$. 
  
In Fig. 3, $\tilde{C}_r^{(\alpha,\infty)}$ (or $\tilde{C}_p^{(\alpha,\infty)}$) 
is plotted as function of the modulus of the 
third component $m$, $-l\leq m \leq l$, of the orbital angular momentum $l$ for different $l$ values 
with a fixed energy, $e_{n,l}=15$, when $\alpha=0.5$ in Fig. 3(a) and $\alpha=2$ 
in Fig. 3(b). It can be observed that $\tilde{C}_r^{(\alpha,\infty)}$ splits again in different sets of  
discrete points. Note that the values associated with 
 the highest $l$ ($l=15$) 
give the minimum values of $\tilde{C}_r^{(\alpha,\infty)}$.   
 
\subsection{Square well}   
   
The eigenstates of the energy in the quantum infinite square well in position space  
for a particle in a box, that is confined in the one-dimensional interval $[0,L]$,  
are given by the wave functions \cite{cohen1977} 
\begin{equation} 
\varphi_k(x)=\sqrt{2\over L}\,\sin\left(k\pi x\over L\right),\,\, k=1,2,\ldots 
\end{equation} 
The probability density of the $kth$ excited state is  
\begin{equation} 
\rho(x)=\mid \varphi_k(x)\mid^2,  
\end{equation} 
that gives a maximum of $2$ when $L$ is considered as the natural length unit in this problem. 
The other factor necessary to obtain the generalized statistical complexity, 
$\tilde{C}_r^{(\alpha,\infty)}$, gives  
\begin{equation} 
e^{R_f^{(\alpha)}}=\left[{2^{\alpha}\over \pi}\int_0^{\pi}\sin^{2\alpha}tdt\;\right]^{1\over 1-\alpha}\,. 
\label{eq-rf} 
\end{equation}  

Then, $\tilde{C}_r^{(\alpha,\infty)}=2g(\alpha)$ with $g(\alpha)=e^{R_f^{(\alpha)}}$.
We conclude that the statistical complexity is degenerated
for all the energy eigenstates of the quantum infinite square well. 
Its value can be computed as a function of $\alpha$. It takes $2$ for $\alpha=0$ and decays
monotonically to $1$ when $\alpha\rightarrow\infty$. 
In the general case of a particle in a $d$-dimensional box of width $L$ in each dimension,
it can be also verified that complexity is degenerated for all its energy eigenstates with a 
constant value given by $\tilde{C}_r^{(\alpha,\infty)}=\left(2\;g(\alpha)\right)^d$.

\section{Summary}  
  
The  generalized complexity measure defined here provides a family of complexity measures. 
We have performed a detailed mathematical characterization of its properties. 
As usual, these complexities have been defined by multiplying two factors, each one bringing global
information on the probability distribution defining the system. 
The whole family is identified by two parameters, $\alpha$ and $\beta$. 
For the special case of $\beta$ going to infinity, it is remarkable that
the new complexity measure is the product of a global quantity by a local information 
of the density distribution.
Then we have carried out the calculation of $\tilde{C}_f^{(\alpha,\infty)}$ 
for different quantum systems: H-atom, harmonic oscillator and square well.
We have found that the behavior of the complexity for these systems in this specific case of global/local 
terms product is similar to that displayed by it in the general case of global/global terms product.

\section*{Acknowledgements}

 \'AN acknowledges grant OTKA No. T\,67923.  ER acknowledges the Spanish project\ 
 FQM-2725 (Junta de Andaluc\'{\i}a) and to FIS2008-01143.


\newpage   
\begin{figure}[h]   
\centerline{\includegraphics[width=8cm]{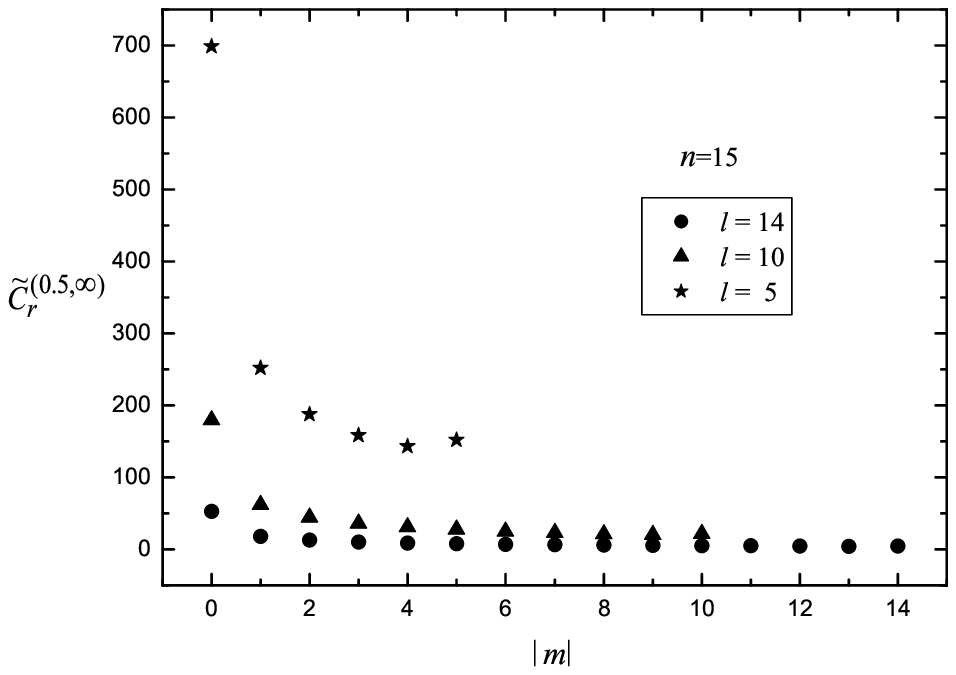}\hskip 5mm\includegraphics[width=8cm]{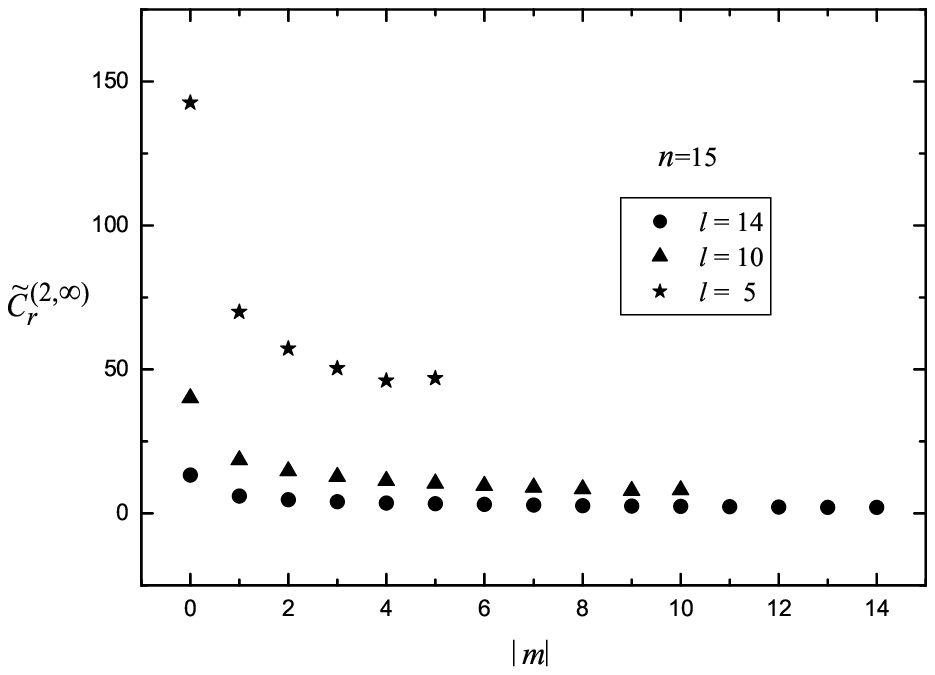}}   
\centerline{(a)\hskip 7cm (b)}    
\caption{Generalized statistical complexity in position space, $\tilde{C}_{r}^{(\alpha,\infty)}$, vs.    
$|m|$ for different $l$ values when $n=15$ in the hydrogen atom. (a) $\alpha=0.5$ and   
(b) $\alpha=2$. All values are in atomic units.}   
\label{fig1}   
\end{figure}

\newpage   
\begin{figure}[h]   
\centerline{\includegraphics[width=8cm]{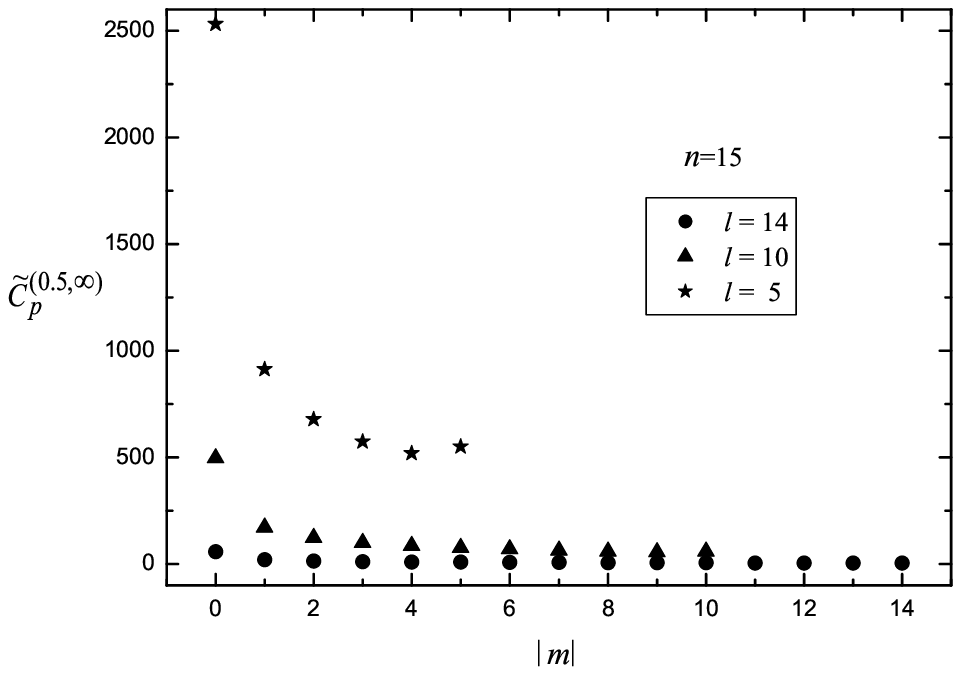}\hskip 5mm\includegraphics[width=8cm]{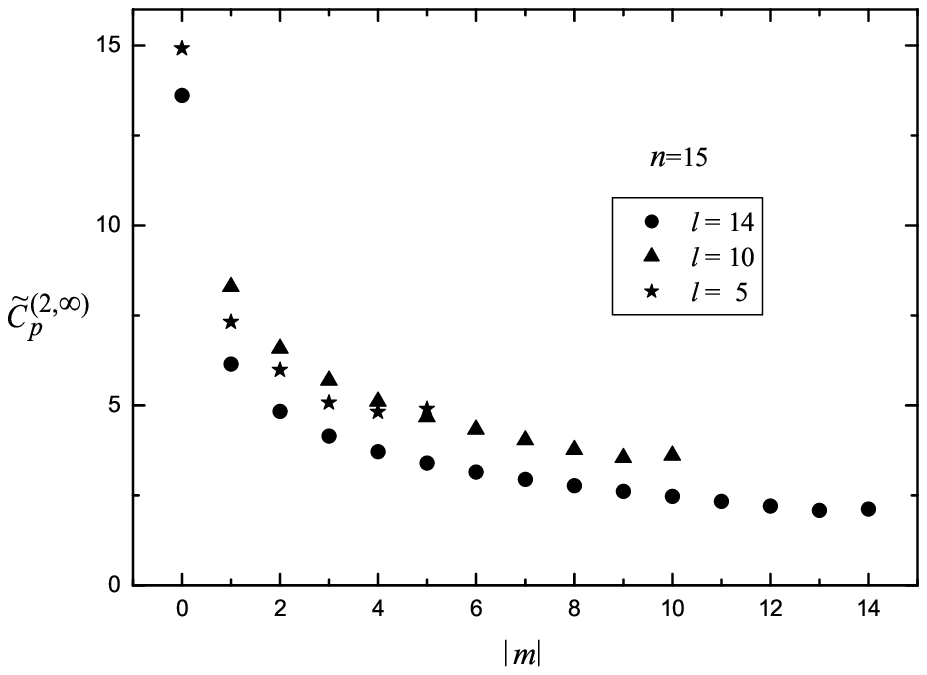}}   
\centerline{(a)\hskip 7cm (b)}    
\caption{Generalized statistical complexity in momentum space, $\tilde{C}_{p}^{(\alpha,\infty)}$, vs.    
$|m|$ for different $l$ values when $n=15$ in the hydrogen atom. (a) $\alpha=0.5$ and $\alpha=2$.  
All values are in atomic units.}   
\label{fig2}   
\end{figure}

\newpage   
\begin{figure}[h]   
\centerline{\includegraphics[width=8cm]{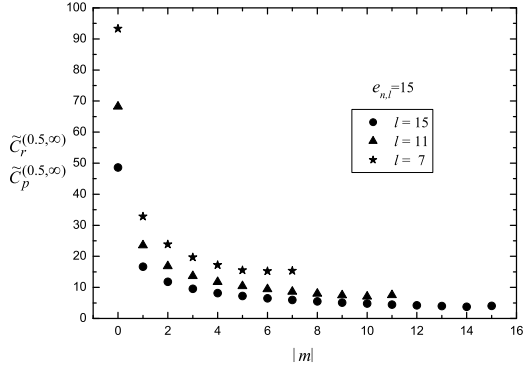}\hskip 5mm\includegraphics[width=8cm]{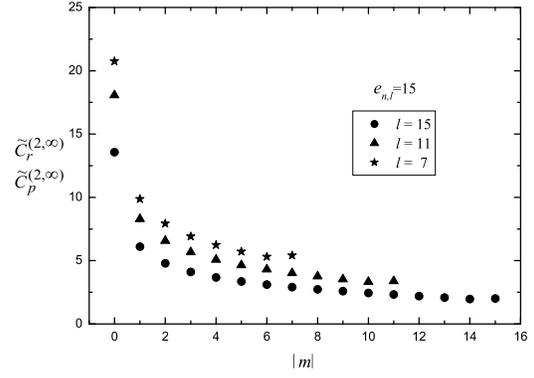}}   
\centerline{(a)\hskip 7cm (b)}    
\caption{Generalized statistical complexity in position space, $\tilde{C}_r^{(\alpha,\infty)}$,  
and momentum space, $\tilde{C}_p^{(\alpha,\infty)}$, vs.  
$|m|$ for the energy $e_{n,l}=15$ in the quantum isotropic harmonic oscillator  
for (a) $\alpha=0.5$ and (b) $\alpha=2$. Recall that $\tilde{C}_r^{(\alpha,\infty)}= 
\tilde{C}_p^{(\alpha,\infty)}$. All values are in atomic units.}   
\label{fig3}   
\end{figure}

\bigskip

\end{document}